# Logical Consistency as a Dynamical Invariant:
# A Quantum Model of Self-Reference and Paradox

N. Cheimarios[1,*,**] and S. Cheimariou[2]

[1]School of Chemical Engineering, National Technical University of Athens Iroon Polytechneiou 9, Zografou 157 72, Athens, Greece

[2] Department of Speech Pathology, Xavier University of Louisiana, 1 Drexel Dr, New Orleans, New Orleans, Louisiana, USA

*Corresponding author: <nixeimar@chemeng.ntua.gr>

**Current address: Accenture, 1 Arcadias 14564 Kifissia, Athens, Greece

**Abstract**

Logical paradoxes and inconsistent information pose deep challenges in epistemology and the philosophy of logic. Classical systems typically handle contradictions only through external checks or by altering the logical framework, as in Tarski's hierarchy or paraconsistent logics. We propose a novel approach: a quantum circuit architecture that intrinsically enforces logical consistency during its unitary evolution. By encoding self-referential or contradictory propositions into a quantum state, the circuit uses interference effects to suppress inconsistent outcomes while preserving coherent ones. We demonstrate this with the Liar Paradox ("This statement is false"), showing that the quantum model naturally stabilizes truth values that would be paradoxical classically. The framework builds on orthomodular quantum logic treating logical propositions as subspace projectors and connects to belief revision and cognitive modeling by providing a physical mechanism for coherence restoration in epistemic states. This work bridges formal logic and quantum computation, suggesting that consistency can be embedded as a structural property of reasoning systems rather than imposed externally.

1. **Introduction**

Human reasoning and formal logical systems both struggle with contradiction. A classic example is the Liar Paradox, the sentence "This statement is false," which cannot be assigned a consistent truth value in a classical two-valued logic [1]. If the sentence is true, then it must be false as it asserts; conversely, if it is false, then it must be true, creating a self-contradictory loop. Such paradoxes reveal fundamental tensions in our conception of truth and rational belief. They challenge the principle of non-contradiction (i.e., a proposition and its negation cannot both be true at the same time) and the assumption that every proposition must be either true or false (principle of bivalence). Philosophers and logicians have long devised strategies to address these issues and they divide broadly into two camps: classical resolutions, which aim to preserve traditional logic by restricting the kinds of statements or truth assignments permitted, and non-classical strategies, which revise foundational principles such as the law of excluded middle or the principle of explosion to accommodate self-reference and inconsistency within the system itself.

For example, within the classical resolutions of the Liar Paradox, Tarski avoided semantic self-reference by introducing a hierarchy of languages, i.e., a *metalogical* stratification that prevents a sentence from asserting its own truth or falsity [2]. In Tarski's hierarchy, truth must always be defined in a higher-level metalanguage. This preserves consistency at the expense of a universal truth predicate. Kripke, in turn, proposed a partial *fixed-point* theory of truth in which the semantic evaluation allows for truth-value gaps: certain sentences (like the liar) remain "ungrounded," neither true nor false [3]. This approach uses a three-valued logic, extending Kleene's K3 logic [4], to achieve a stable, consistent assignment of truth values via iterative approximation. Related revision theories, Feferman, and Gupta & Belnap's revision semantics [5–8], go further, allowing truth values to oscillate through successive approximations until a consistent *limiting state* (if one exists) is reached [8]. These approaches illustrate a common theme: when faced with paradox or inconsistency, classical frameworks often enforce consistency by moving outside the original system, whether through metalanguages, multi-valued logics, or iterative procedures.

In the non-classical strategies, a general response it to weaken or modify the logic itself. For example, in paraconsistent logics [9,10], contradictions do not entail an explosion of triviality. In classical logic, accepting both a proposition A and its negation ¬A allows any proposition whatsoever to be derived. This collapse is known as the principle of explosion (ex falso quodlibet). This leads to triviality, where every statement becomes provable and the system loses its discriminatory power. Paraconsistent logics avoid this outcome by rejecting the inference from contradiction to arbitrary conclusions. A paraconsistent consequence relation "accommodates inconsistency in a controlled way", allowing contradictory statements to coexist without making every statement true, i.e., without trivialization [11]. This is one way to formalize reasoning "inconsistency-tolerantly", so that a knowledge base can contain P and ¬P yet not infer arbitrary Q. Such logics (e.g. dialetheism, which accepts true contradictions, or relevant logics, which restrict inference rules) provide alternatives to classical reasoning under contradiction. Meanwhile, the field of belief revision addresses how a rational agent should modify its belief set when new information creates inconsistency. The influential AGM framework [12] axiomatically characterizes how to minimally retract or adjust beliefs to restore consistency after a contradiction is introduced. In essence, belief revision theories treat consistency as an important post facto requirement: if a contradiction arises, some belief must be weakened or given up to preserve an overall coherent belief state.

Even outside formal logic, the need to reconcile contradictions is evident. In cognitive psychology, cognitive dissonance [13] occurs when an individual holds inconsistent beliefs or attitudes, often resulting in psychological pressure to resolve the inconsistency. The mind's resolution strategies (e.g. adjusting beliefs or adding rationalizations) can be seen as a dynamical process of regaining coherence in one's worldview. In artificial intelligence, truth maintenance systems, e.g. Doyle's TMS [14,15], explicitly track dependencies and retract assumptions to maintain consistency in a knowledge base. These various perspectives, philosophical, logical, psychological, and computational, all converge on the notion that logical consistency is a fundamental hallmark of rationality and coherent reasoning. Yet, in classical settings, consistency is typically enforced by external interventions: meta-theoretical constraints, manual checks, or non-standard logic formalisms.

In this work, we explore a different approach inspired by the formalism of quantum mechanics. We ask: Can the structure of a physical system itself enforce logical consistency, without external oversight? Our proposal leverages the properties of quantum computation, in particular, the phenomenon of interference in unitary quantum circuits, to embed logical consistency within the evolution of a system. The key idea is to map logical propositions and their potential contradictions to quantum states such that any inconsistent combination of truth values cancels itself out by destructive interference, while consistent combinations reinforce and persist. In other words, we construct a quantum-mechanical filter that only allows logically coherent states to survive as measurable outcomes. Crucially, this consistency enforcement is not an external check but a direct consequence of the system's unitary dynamics.

Our approach draws on the heritage of quantum logic in the sense of Birkhoff and von Neumann [16]. Quantum logic, as an orthomodular lattice of propositions, was originally proposed as a non-classical logic that better aligns with quantum phenomena [17]. We take a step further by providing a concrete operational realization of a quantum-logical consistency constraint. In our circuits, a logical proposition corresponds to a projector onto a subspace of the Hilbert space, as in the quantum logic formalism [16]. The condition of "all constraints satisfied" or "no contradiction present" is itself represented by a projector (a measurement or subspace) that the unitary evolution reflects onto. As a result, the unitary circuit acts as a logical predicate test: states that lie in the "consistent" subspace evolve trivially (identity action), whereas states with any inconsistency component acquire a phase kick that causes them to interfere destructively with their coherent counterparts. This operationalization of a logical requirement has no classical analog in standard Boolean circuits, which would require an external agent to check and prune inconsistent assignments.

By encoding logical relations into quantum entanglement and interference, our work connects to recent interdisciplinary trends. In quantum cognition, researchers have used quantum probability and vector spaces to model human concept combinations and decision paradoxes that defy classical logic or probability theory [18–21]. These studies highlight that human reasoning may follow non-classical patterns, for example, showing interference effects in decision outcomes, and quantum formalisms can capture such patterns. Our approach contributes to this line of thought by providing a tangible quantum model of reasoning under contradiction. It suggests that the mind's resolution of a paradox or inconsistent beliefs could be viewed analogously to a quantum system finding a stable superposition that neutralizes the inconsistency. Indeed, as we show through the Liar Paradox example, the quantum circuit's final state embodies a kind of psychological equilibrium between asserting and denying a self-referential statement, i.e, a superposed "truth state" that avoids outright contradiction by not settling on a classical truth value.

In summary, our contribution is a unitary quantum circuit architecture that intrinsically enforces logical consistency and dynamically manages paradoxical configurations. We demonstrate this by constructing a quantum circuit for the Liar Paradox and showing how interference suppresses inconsistent configurations while preserving coherent outcomes. We then generalize the approach and discuss its significance: (i) as a formal method linking logical constraints to physical dynamics, (ii) as a novel perspective on truth stabilization drawing parallels to semantic fixed-point and revision-based theories, but realized here through unitary evolution rather than meta-logical rules, and (iii) as a framework for modeling coherence in reasoning systems without assuming classical truth assignments. The central contribution of this work is to show that logical coherence can be implemented as a dynamical invariant of unitary evolution, rather than as a meta-logical constraint imposed on a formal system.

## 2. Quantum Circuits as Consistency Enforcers

In this section we describe how a quantum circuit can represent a simple logical system that might become inconsistent, and how unitary evolution can dynamically enforce logical coherence. We first develop an operator framework in terms of projectors and reflections. We then show how this framework is realized in a concrete four-qubit circuit for the Liar Paradox and how the same architecture scales to larger systems. Finally, we present simulation and hardware results.

### 2.1. Operator framework: projectors, reflections, and coherence flags

Let the total Hilbert space for $N$ qubits be

$$\Pi = \otimes_{k=1}^{N} \mathbb{C}^2 \tag{1}$$

of dimension $2^N$. Each logical component acts on a subset of this tensor product.

For each *statement–negation* pair $(q_{c_i}, q_{r_i})$, define the local **contradiction projector**

$$P_i = |q_{c_i} = 1, q_{r_i} = 0\rangle\langle q_{c_i} = 1, q_{r_i} = 0| = |10\rangle_{(q_{c_i}q_{r_i})}\langle 01| \tag{2}$$

This operator acts on the two-qubit subspace $\mathcal{H}_{(c_i r_i)} \subset \mathcal{H}$ and annihilates all logically consistent configurations. The complementary consistency projector is,

$$\Pi_i = I_{(q_{c_i}q_{r_i})} - P_i \tag{3}$$

which projects onto the subspace of consistent logical relations. Its image $\mathrm{Im}(\Pi_i)$ therefore defines the local *consistency manifold* for that pair. The global consistency projector is given by the tensor product,

$$\Pi = \otimes_{i=1}^{N/2} \Pi_i \tag{4}$$

acting on the full space $\mathcal{H}$. It satisfies $\Pi^2 = \Pi = \Pi^\dagger$, and its image $\mathrm{Im}(\Pi)$ defines the globally coherent subspace.

The logical-coherence circuit implements the Hermitian and unitary operator

$$U = 2\Pi - I \tag{5}$$

which acts as a **reflection** about the consistent subspace. For any state $|\psi\rangle \in \mathcal{H}$,

$$U|\psi\rangle = \begin{cases} |\psi\rangle, & \text{if } \Pi|\psi\rangle = |\psi\rangle \\ -|\psi\rangle, & \text{if } \Pi|\psi\rangle = 0 \end{cases} \tag{6}$$

Hence $U^2 = I$ and the eigenvalues $\{+1, -1\}$. Operationally, the full as a block composition of local reflections sharing a global *coherence flag* $q_f$,

$$U = \bigotimes_{i=1}^{N/2} \left(2\Pi_i - I_{(q_{c_i} q_{r_i})}\right) \otimes I_{q_f} \qquad (7)$$

$q_f$ serves as a measurable indicator of logical consistency. Define the Pauli-ZZZ operator on that qubit,

$$Z_f = I^{\otimes(N-1)} \otimes Z \qquad (8)$$

which acts as $Z_f|0\rangle = |0\rangle, Z_f|1\rangle = -|1\rangle$. The corresponding coherence observable is the expectation value,

$$\langle Z_f \rangle = \langle \psi | F | \psi \rangle = P_f(0) - P_f(1) \qquad (9)$$

Here $P_f(0)$ and $P_f(1)$ denote the measured probabilities of the flag being 0 or 1.

A value $\langle Z_f \rangle \approx 1$ indicates full logical consistency (coherence preserved), while $\langle Z_f \rangle \approx -1$ corresponds to maximal inconsistency (phase inversion due to contradiction).

Define the logical Hamiltonian

$$H_{logic} = I - \Pi = \sum_{i=1}^{N/2} P_i \qquad (10)$$

which penalizes inconsistencies analogously to QCS projectors. The unitary operator can then be expressed as

$$U = e^{-iH_{logic}\pi} \qquad (11)$$

since $e^{-i\pi P} = I - 2P$ for any projector $P$. See Supplementary Material (SM) Section 2 (S2) for the derivation of Eq. (11) and the proof that the unitary operator constitutes a reflection about the consistent subspace. Therefore, the fixed-point condition,

$$U|\psi\rangle = |\psi\rangle \Leftrightarrow H_{logic}|\psi\rangle = 0 \tag{12}$$

establishes that the invariant subspace of the unitary evolution coincides with the kernel of the logical Hamiltonian. Logical coherence is thus dynamically enforced by interference rather than by energy minimization.

In standard quantum constraint satisfaction (QCS) or quantum satisfiability (QSAT) formulations, logical consistency is defined *statically* in terms of an energy landscape. A problem instance is represented by a constraint Hamiltonian

$$H = \sum_i P_i \tag{13}$$

where each $P_i$ is a projector penalizing violation of the *i*-th logical constraint. The ground-state manifold of $H$ then corresponds to the set of satisfying assignments—that is, logically consistent states. In these models, consistency arises through energetic minimization or projector constraint satisfaction: the system's dynamics are used to find the minimum-energy state, but logical coherence itself is not a dynamical property of evolution. By contrast, the present model employs Eq. (11), making logical coherence a *conserved quantity* of reversible evolution rather than a variational minimum. Thus, the approach presented here defines logical coherence as a property of unitary evolution itself, not as an outcome of minimization. States representing logical contradictions acquire destructive interference phases and are dynamically canceled, while logically admissible states remain invariant under $U$. Hence, coherence is enforced by interference rather than simulated by energetics.

Let the first $n$ qubits $(q_{c_1}, \ldots, q_{c_n})$ encode potential contradictions, and the next $n$ qubits $(q_{r_1}, \ldots, q_{r_n})$ represent available resolution mechanisms. The circuit applies a sequence of Toffoli-style gates, that flips $q_f$ only when $q_{c_i} = 1$ and $q_{r_i} = 0$. The effective rule can be expressed as,

$$f_{out} = f_{in} \oplus \bigvee_{i=0}^{n} \left( q_{c_i} \wedge \neg q_{r_i} \right) \qquad (14)$$

The coherence flag $q_f$ starts in the coherent state $f_{in}=1$. Each contradiction–resolution pair $(q_{c_i}, q_{r_i})$ is tested; if a contradiction $q_{c_i}$, is active and remains unresolved ($q_{r_i} = 0$) the flag is flipped once. If no such contradictions exist, the flag remains 1, signaling global logical consistency – see also Table 1. A detailed proof that the unitary operator acts as a fixed-point reflection about the consistent subspace is provided in the SM, S1.

**Table 1**: Logical behavior of the model

| Contradictions ($q_{c_i}$) | Resolutions ($q_{r_i}$) | Output flag ($f_{out}$) | Interpretation |
|---|---|---|---|
| None active | Any | 1 | Fully consistent |
| Some active but all resolved | Any other resolutions | 1 | Locally resolved |
| At least one unresolved contradiction | - | 0 | Inconsistency detected |
| All contradictions unresolved | - | 0 | Fully inconsistent |

Equations (1)–(14) define the complete operator framework of the logical-coherence circuit. The operator $U$ acts as a reflection symmetry about the consistent subspace, $H_{logic}$ formalizes the underlying constraint structure, and $Z_f$ provides a measurable physical observable of logical coherence. Together, these operators form a self-consistent algebra linking logical projection, unitary dynamics, and observable coherence, in full accordance with the standard operator formalism of quantum computation [22].

## 3. Application: The Liar Paradox

### 3.1 Logical roles of the qubits

For the concrete Liar circuit (Fig. 1), we use a four-qubit register:

- $q_0$: statement qubit, encoding the truth of "this statement is true/false" (the content of the liar sentence).
- $q_1$: negation tracker, encoding "this statement is false".
- $q_2$: resolution qubit, acting as a local contradiction detector.
- $q_3$: global coherence flag, signaling whether paradoxical superpositions have been suppressed.

The initial configuration $|0000\rangle$ corresponds to a self-consistent baseline state prior to evaluation.

Intuitively:

- The pair $(q_0, q_1)$ plays the role of a specific $(q_{c_i}, q_{r_i})$ pair from the operator formalism, encoding the relationship between a statement and its negation.
- The resolution qubit $q_2$ indicates when the Liar configuration is paradoxical.
- The global flag $q_3$ records whether paradoxical branches survive or are suppressed by interference.

### 3.2 The quantum circuit for the Liar Paradox

The quantum circuit of Fig.1 models the Liar Paradox form "This statement is false." If the statement ($q_0 = 1$) implies its own negation ($q_1 = 1$) without an accompanying resolution ($q_2 = 0$), coherence is violated, and destructive interference removes this branch from the system's final state.

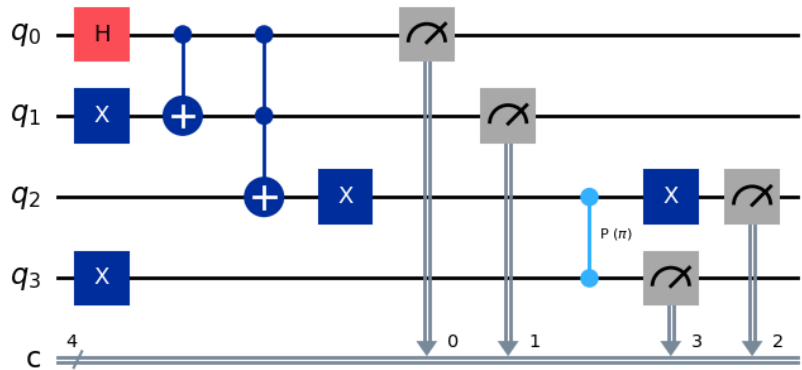

**Figure 1**. The 4-qubit circuit for resolving the Liar Paradox.

The circuit begins by applying a Hadamard gate to the statement qubit $q_0$, preparing it in a coherent superposition of truth and falsehood. This represents the inherent semantic ambiguity of the self-referential sentence, transforming $|0\rangle$ into $\frac{|0\rangle+|1\rangle}{2}$. Next, a controlled-NOT operation is applied from $q_0$ to $q_1$, coupling the truth value of the statement to its own negation tracker such that self-reference is explicitly encoded in the joint quantum state. A subsequent Toffoli gate with controls on $q_0$ and $q_1$ flips the resolution qubit $q_2$ when both the statement and its negation simultaneously evaluate to true, thereby detecting the paradoxical configuration ($q_0 = 1$, $q_1 = 1$).

Logical coherence is then enforced through a controlled phase operation. Specifically, a π-phase shift is conditionally applied to the global consistency qubit $q_3$ when the paradox has not been resolved, i.e., when $q_2 = 0$. This introduces a relative sign difference between consistent and inconsistent logical branches, producing destructive interference for paradox-supporting amplitudes in the final statevector.

Taken together, these steps define a unitary evolution $UUU$ acting on the four-qubit system as:

$$U = CP_{q_2 \to q_3}(\pi) \cdot CCX_{q_0, q_1 \to q_2} \cdot CNOT_{q_0 \to q_1} \cdot (H \otimes I \otimes I \otimes I) \tag{15}$$

This sequence realizes a physically grounded mechanism by which self-contradictory logical assignments suppress their own contributions to the final measurement statistics, leaving only consistent truth valuations observable.

The proposed logical-coherence architecture can be extended beyond the four-qubit prototype to arbitrary *N*-qubit systems while retaining a compact and scalable structure. Each *statement–negation* pair contributes one contradiction qubit ($q_c$) and one resolution qubit ($q_r$), and all pairs share a single global coherence flag ($q_f$) that monitors consistency across the system. The corresponding unitary operator can be expressed as a product of local conditional operations,

$$U_N = \prod_{i=1}^{N/2} CCX(q_{c_i}, q_{r_i}, q_f) \, CP(\pi, q_f, q_{r_i}) \tag{16}$$

which generalizes the 4-qubit kernel to any even *N*. Each contradiction–resolution pair activates one Toffoli (CCX) and one controlled-phase (CP) gate acting on the shared coherence qubit. Details about the asymptotic complexity, connectivity constraints and error accumulation can be found the SM.

### 3.3 Feasibility on current hardware

Because both the two-qubit gate count and circuit depth scale linearly with the number of qubits *N*, the proposed architecture remains compatible with current mid-scale IBM Quantum hardware, such as the 27- and 65-qubit *Falcon* and *Hummingbird* processors. For a representative case of $N = 8$, the compiled circuit is expected to contain approximately 200–250 two-qubit operations, which approaches—but does not exceed—the typical coherence window of present superconducting qubit systems. Contemporary IBM devices report median relaxation times $T_1$ in the range of 80 –150 µs and two-qubit gate error rates around $10^{-3}$, with single-qubit errors typically below $10^{-4}$ [29]. Under these conditions, the cumulative gate infidelity and decoherence effects remain tolerable for circuits of this depth, supporting feasibility for eight-qubit demonstrations. Furthermore, employing relative-phase Toffoli decompositions or incorporating mid-circuit measurement and reset operations can reduce the effective two-qubit-gate depth by roughly 50%, providing additional margin within the NISQ-era coherence constraints.

## 4. Results: Ideal simulations and hardware executions

The logical-coherence circuit is executed both on an ideal noiseless simulator and on IBM's ibm_torino quantum processor. The analysis begins with an idealized 4-qubit quantum circuit implementing the Liar Paradox. The circuit's final state vector, obtained from a noiseless simulation, provides a complete representation of the logical system's internal dynamics without any hardware imperfections. In this ideal setting, all quantum gates are assumed to operate with perfect fidelity, and decoherence, crosstalk, and readout noise are entirely absent. This permits the pure logical structure of self-reference and contradiction to emerge directly from unitary evolution

The four-qubit register represents distinct logical roles: the first qubit ($q_0$) encodes the proposition "this statement is true," the second ($q_1$) tracks its negation ("this statement is false"), the third ($q_2$) acts as a resolution or contradiction detector, and the fourth ($q_3$) serves as a global coherence flag that signals when paradoxical superpositions are suppressed. The initial configuration |0000⟩ corresponds to a self-consistent state prior to evaluation.

The initial configuration |0000⟩ corresponds to a self-consistent state prior to evaluation. Under ideal, noise-free evolution, the final state vector contains only two nonzero amplitudes, corresponding to the basis states |1010⟩ and |1001⟩, each with probability 0.5 (Fig. 2). All other computational basis states are completely excluded because they violate the logical constraints embedded in the circuit. The state |1010⟩ represents the configuration in which the statement asserts its own falsity — the "liar" condition — while the coherence flag ($q_3 = 0$) indicates that this contradiction is detected but remains unresolved.

The complementary state |1001⟩ captures the self-referential resolution of the paradox, where the circuit restores global coherence ($q_3 = 1$) through destructive interference of inconsistent amplitudes. The total exclusion of all other states such as |0000⟩, |0101⟩, or |1110⟩ demonstrates the circuit's declarative filtering behavior: only logically admissible configurations survive, while contradictory branches are physically inaccessible in the ideal

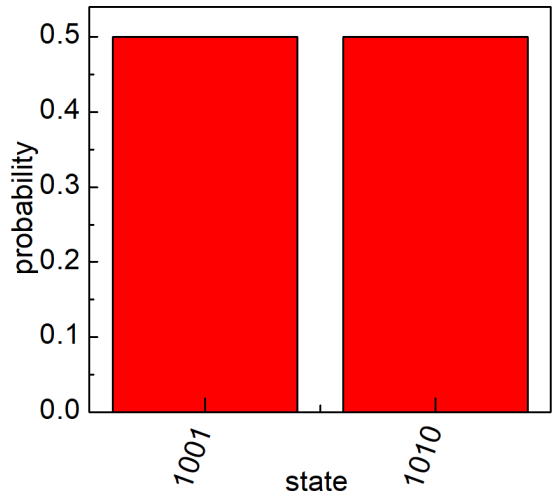

**Figure 2.** Measurement outcomes from ideal simulations (state-vector computation). Only the logically valid states |1001⟩ and |1010⟩ appear, each with probability 0.5. All other states have zero probability and are omitted for clarity.

To validate this logical mechanism under realistic conditions, the same 4-qubit circuit is executed on IBM's *ibm_torino* quantum processor [30–32]. Despite the presence of hardware noise, gate imperfections, and readout errors [33], the experimental probability distribution (Fig. 3) preserved the structure of the ideal case. The two logically admissible configurations, |1010⟩ and |1001⟩, remained dominant in the measurement results, while all other states appear only as minor noise-induced components.

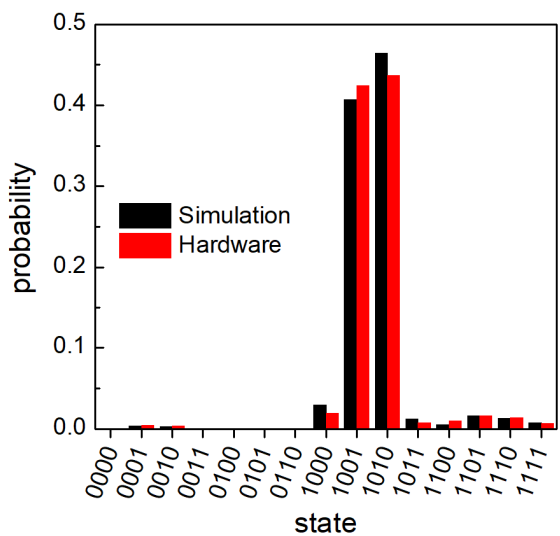

**Figure 3.** Measurement outcomes from executions on IBM's *ibm_torino* quantum processor (hardware and simulation). Probabilities are averaged over 20 independent trials. The dominant states |1010⟩ and |1001⟩ confirm that the Liar-Paradox circuit maintains its logical structure under real-hardware conditions.

To quantitatively assess the logical-coherence performance of the proposed circuit, four complementary metrics are computed for both the ideal (simulated) and hardware executions. Each metric is formally defined below and evaluated as the mean ± standard deviation over twenty independent runs of the experiment on IBM's *ibm_torino* backend, each consisting of 8192 shots. See SM S3 for detailed probability distributions and raw measurement outcomes used in Fig. 3.

To quantify logical coherence, we define a consistency-fidelity metric i.e. the total probability mass that falls within logically consistent outcomes, as,

$$F_C = \sum_{x \in S_{cons}} p_{exp}(x) \qquad (17)$$

where $x$ represents each possible outcome in the probability space, $S_{cons}$ is the set of bitstrings that represent *logically consistent* outcomes and $p_{exp}(x)$ the probability observed in the experiment for that bitstring. $F_C$ measures how much of the experiment's observed probability distribution lies within the consistent region. If $F_C = 1$, all outcomes are consistent while if $F_C < 1$, then some experimental weight falls on inconsistent outcomes, implying imperfections, decoherence, or logical violations.

The total variation distance between an experimental distribution $p_{exp}(x)$ and an ideal (or simulated) distribution $p_{ideal}(x)$ is defined as,

$$D_{TV} = \frac{1}{2} \sum_x |p_{exp}(x) - p_{ideal}(x)| \qquad (18)$$

The factor 1/2 ensures that the total variation distance ranges from 0 (identical distributions) to 1 (completely non-overlapping distributions).

To quantifiy how strongly paradoxical outcomes are suppressed we use the interference-suppression ratio defined as,

$$R_I = 1 - \frac{\sum_{x \in S_{paradox}} p_{exp}(x)}{\sum_{x \in S_{paradox}} p_{ideal}(x)} \tag{19}$$

The $\chi^2$ Goodness-of-Fit $(p_{\chi^2})$ test compares observed and expected frequency distributions under the null hypothesis $H_0: p_{hardware} = p_{real}$. The resulting *p*-value quantifies statistical agreement between the two distributions.

Quantitative comparison of the experimental and ideal results (Table 2) demonstrates that coherence is robustly preserved under realistic hardware noise. The consistency fidelity, defined as the total probability of all logically valid configurations, is $F_{C(sim)} = 0.904 \pm 0.008$ for the simulator and $F_{C(hardware)} = 0.907 \pm 0.008$ for the hardware, confirming that the overall distribution of consistent outcomes remains virtually unchanged. The total variation distance between the simulated and experimental probability distributions was $D_{TV} = 0.024 \pm 0.013$, corresponding to less than a 3% deviation from the ideal statevector simulation. The interference suppression ratio is $R_i = 2.1\% \pm 9.8\%$, a small and statistically noisy value consistent with the near-complete cancellation of paradox-supporting amplitudes. Because the ideal distribution already assigns negligible probability to inconsistent branches, residual fluctuations in $R_i$ primarily reflect stochastic sampling and readout noise rather than a structural failure of the interference mechanism. Finally, a chi-squared ($\chi^2$) goodness-of-fit test yields a mean $p = 0.10 \pm 0.17$, indicating no statistically significant deviation between simulated and experimental distributions at the 95% confidence level. Together, these results verify that the logical-coherence circuit performs as designed, maintaining high-fidelity paradox suppression and stable logical behavior across repeated hardware executions.

Table 2: Averaged metrics for the Liar Paradox circuit

| | |
|---|---|
| Consistency Fidelity (simulation) | 0.904 ± 0.008 |
| Consistency Fidelity (hardware) | 0.907 ± 0.008 |
| Total Variation Distance | 0.024 ± 0.013 |
| Interference Suppression | 2.1% ± 9.8% |
| $\chi^2$ test p-value | 0.1000.172 |

5. **Discussion**

Here, we presented a quantum circuit model that treats logical coherence not as an externally imposed constraint but as an intrinsic property of unitary evolution. The model shows that contradiction management, typically handled by meta-logical rules, can instead be realized through the physical structure of quantum dynamics. The circuit is not a numerical simulation of logical rules, but an implementation in which logical constraints are encoded directly into the unitary evolution itself. The experimental confirmation of only two stable configurations, $|1010\rangle$ and $|1001\rangle$, illustrates how the quantum formulation of the Liar Paradox settles into a self-referential equilibrium. These outcomes encode complementary but mutually dependent truth-assignments: $|1010\rangle$ represents the classical paradox ("this statement is false") by maintaining truth and falsity in tension, while $|1001\rangle$ represents a stabilized reinterpretation in which destructive interference removes the contradictory amplitude. Rather than producing inconsistency, the evolution stabilizes a superposition of the two admissible interpretations. Importantly, the model does not claim to resolve the Liar Paradox by assigning it a determinate classical truth value. Instead, each measurement yields a single consistent valuation, either true or false, while the underlying quantum state encodes the coexistence of both possibilities. An observer interacting with the system is therefore never forced into accepting a contradiction. This superposed state forms a quantum analogue of semantic oscillation, paradoxical under classical logic but rendered coherent in the quantum regime, where inconsistent amplitudes cancel via interference and consistency emerges as an invariant of the evolution.

The empirical behavior of the circuit supports this interpretation. High consistency fidelity and minimal total-variation distance between ideal and hardware outputs show that paradox-suppressing interference persists even on NISQ devices. The small residual interference-suppression ratio reflects the fact that, in the ideal circuit, contradictory amplitudes approach zero; deviations arise primarily from hardware noise rather than from conceptual failure. These results demonstrate that self-reference can be encoded without yielding contradiction, and that paradox resolution can be approached as a dynamical property of Hilbert-space processes. The circuit thereby implements a fixed point of the truth operator, with the inconsistent components eliminated by the evolution itself.

From a logical perspective, the design functions as a physical analogue of the projectors that define consistent subspaces in Birkhoff–von Neumann quantum logic [16]. Logical consistency arises dynamically: inconsistent configurations interfere destructively, leaving only admissible truth assignments in the output distribution. This yields a physically grounded method for encoding logical constraints into unitary operators and suggests a pathway for quantum implementations of constraint-satisfaction, model-checking, or automated reasoning tasks. Crucially, the coherence condition is not enforced by symbolic rules; it is realized by the physical evolution, making consistency a structural property of the information state.

This framework has clear connections to philosophical accounts of self-reference and truth. Classical treatments of the Liar Paradox [3,8] resolve self-reference by identifying a semantic fixed point within an abstract space of valuations. The quantum circuit we propose achieves an analogous outcome, but in the space of amplitudes: contradictory components are driven to zero, and the surviving superposition behaves as a fixed point of the truth predicate. The resulting state assigns no single classical truth value to the liar sentence, yet guarantees that any observed valuation will be consistent. Truth, on this view, becomes a global property of the epistemic state rather than a locally assigned binary. This aligns naturally with coherence theories of truth, where the acceptability of a proposition depends on its integration within a consistent global structure.

The model also provides a new angle on paraconsistent logic [9,10]. Paraconsistent systems typically block the rule of explosion, allowing P and ¬P without triviality. In our construction, something stronger occurs: the joint state encoding $P \land \neg P$ acquires zero amplitude and is therefore unobservable. Contradictions do not merely fail to explode; they fail to occur. The circuit is not dialetheic, since it never outputs a classical record claiming that a proposition is both true and false; yet it accommodates superposed contradictory components in a non-classical manner. This parallels features of quantum logic where distributivity fails and truth conditions cannot be reduced to classical valuations of propositions. The model thus offers a bridge between paraconsistent reasoning and quantum-theoretic semantics, showing how non-classical truth-behavior can arise from the structural properties of quantum evolution.

The implications for belief revision [12] are similarly structural. In classical approaches, resolving inconsistency requires explicit decisions about which beliefs to retract. In the quantum model, revision is implemented implicitly through interference: incompatible commitments annihilate one another. The circuit does not choose which side of the contradiction to discard; the evolution explores both possibilities in superposition and converges to a stable pair of consistent outcomes. A single measurement yields one coherent interpretation, but the underlying mechanism ensures that no inconsistent interpretation is ever realized. This suggests a dynamical analogue to epistemic update: revision is not performed by rule but emerges from the intrinsic behavior of the system.

Our hardware implementation further demonstrates that the model is more than a theoretical construction. Executed on IBM's *ibm_torino* device, the circuit reliably produced the two admissible outcomes with high probability, and inconsistent outcomes appeared only at noise-level frequencies. This provides empirical evidence that quantum dynamics can physically enforce logical coherence. The device was not programmed to suppress contradictions; the unitary structure ensured that inconsistent amplitudes canceled, showing how logical constraints can be instantiated directly in physical processes.

Beyond its role in modeling self-referential paradoxes, the proposed logical-coherence circuit provides a general mechanism for enforcing global consistency across quantum systems subject to local logical constraints. Acting as a unitary logical filter, it suppresses amplitudes corresponding to inconsistent configurations while preserving the coherent subspace of valid logical assignments.

This capability has broad implications for several fields: in quantum optimization, it can function as a dynamic constraint-satisfaction operator analogous to QAOA, improving convergence toward feasible solutions [34,35]; in neuro-symbolic and hybrid AI systems, it may serve as a quantum layer that enforces logical consistency within superposed reasoning states [36]; in quantum hardware, its interference-based structure offers a reversible alternative to projective error correction and decoherence control [33]; in quantum linguistics and cognitive modeling, it captures semantic coherence in context-dependent or self- referential statements [18,20] and in distributed decision-making and game theory, it provides a physical basis for enforcing Nash-consistent or Pareto-optimal strategies through interference [37,38].

More broadly, viewing logical coherence as a conserved quantity within a dynamical system offers a novel perspective on the relation between logic and physics. It suggests that consistency may function analogously to a physical invariant: not something checked after the fact, but something preserved by lawful evolution. This stance resonates with dynamic epistemic logic, which emphasizes the process of updating information states, while adding the suggestion that physical dynamics themselves may instantiate epistemic transition rules.

## 6. Conclusions

Embedding logical consistency within unitary quantum circuits creates new opportunities for dialogue between logic, quantum theory, and the philosophy of information. It indicates that the resolution of paradox and self-reference can be achieved not only through conceptual reformulation but also through physically grounded principles such as superposition and interference. This raises further questions of philosophical and technical relevance: which other logical or epistemic norms can be modeled as invariants of physical evolution? Can semantic paradoxes be experimentally investigated through quantum implementations? And might insights from theories of truth and reasoning guide the design of quantum algorithms for knowledge representation and constraint management? The present work suggests that logical coherence can emerge as an intrinsic property of a physical process, and that integrating epistemic logic with quantum computation may yield both conceptual clarity and new computational tools.

**References**

1. Beall J, Glanzberg M, Ripley E. Liar Paradox. (2011) Available at: https://plato.stanford.edu/archives/fall2025/entries/liar-paradox/ [Accessed November 21, 2025]

2. Tarski A. *Logic, Semantics, Metamathematics: Papers from 1923 to 1938*. Hackett Publishing (1983).

3. Kripke S. Outline of a Theory of Truth. *The Journal of Philosophy* (1976) **72**:690–716. doi:10.2307/2024634

4. Fitting M. Kleens's three valued logics and their children. *Fundamenta Informaticae* (1994) **20**:113–131. doi:10.3233/FI-1994-201234

5. Belnap ND. Gupta's Rule of Revision Theory of Truth. *Journal of Philosophical Logic* (1982) **11**:103–116. doi:10.1007/bf00302340

6. Feferman S. Toward useful type-free theories. I. *Journal of Symbolic Logic* (1984) **49**:75–111. doi:10.2307/2274093

7. Gupta A. Truth and Paradox. *Journal of Philosophical Logic* (1982) **11**:1–60.

8. Gupta A, Belnap ND. *The Revision Theory of Truth*. MIT Press (1993).

9. Priest G. "Paraconsistent Logic," in *Handbook of Philosophical Logic*, eds. D. M. Gabbay, F. Guenthner (Dordrecht: Springer Netherlands), 287–393. doi:10.1007/978-94-017-0460-1_4

10. Arruda AI, Asenjo FC, Batens D, Brady R, Bunder M, Costa NCA da, Giambrone S, Goldstein L, Jennings R, Kotas J, et al. *Paraconsistent Logic, Essay on the Inconsistent*. Philosophia (1989). doi:10.2307/j.ctv2x8v8c7

11. Priest G. *In Contradiction*. Oxford University Press (2006). doi:10.1093/acprof:oso/9780199263301.001.0001

12. Alchourrón CE, Gärdenfors P, Makinson D. On the logic of theory change: Partial meet contraction and revision functions. *The Journal of Symbolic Logic* (1985) **50**:510–530. doi:10.2307/2274239

13. Miller MK, Clark JD, Jehle A. "Cognitive Dissonance Theory (Festinger)," in *The Blackwell Encyclopedia of Sociology* (John Wiley & Sons, Ltd). doi:10.1002/9781405165518.wbeosc058.pub2

14. Doyle J. A truth maintenance system. *Artificial Intelligence* (1979) **12**:231–272. doi:10.1016/0004-3702(79)90008-0

15. Doyle J. "Mechanics and Mental Change," in *Evolution of Semantic Systems*, eds. B.-O. Küppers, U. Hahn, S. Artmann (Berlin, Heidelberg: Springer), 127–150. doi:10.1007/978-3-642-34997-3_7

16. Birkhoff G, Von Neumann J. The Logic of Quantum Mechanics. *Annals of Mathematics* (1936) **37**:823–843. doi:10.2307/1968621

17. Jauch JM. *Foundations of Quantum Mechanics*. Addison-Wesley Publishing Company (1968). Available at: https://books.google.gr/books?id=FwpRAAAAMAAJ

18. Aerts D, Sozzo S, Veloz T. New fundamental evidence of non-classical structure in the combination of natural concepts. *Philosophical Transactions of the Royal Society A: Mathematical, Physical and Engineering Sciences* (2016) **374**:20150095. doi:10.1098/rsta.2015.0095

19. Aerts D. Quantum structure in cognition. *Journal of Mathematical Psychology* (2009) **53**:314–348. doi:10.1016/j.jmp.2009.04.005

20. Bruza PD, Fell L, Hoyte P, Dehdashti S, Obeid A, Gibson A, Moreira C. Contextuality and context-sensitivity in probabilistic models of cognition. *Cognitive Psychology* (2023) **140**:101529. doi:10.1016/j.cogpsych.2022.101529

21. Busemeyer JR, Bruza PD. *Quantum Models of Cognition and Decision*. Cambridge: Cambridge University Press (2012). doi:10.1017/CBO9780511997716

22. Nielsen MA, Chuang IL. *Quantum Computation and Quantum Information: 10th Anniversary Edition*. Anniversary edition. Cambridge ; New York: Cambridge University Press (2011).

23. Yuan P, Allcock J, Zhang S. Does qubit connectivity impact quantum circuit complexity? (2023) doi:10.48550/arXiv.2211.05413

24. Sim S, Johnson PD, Aspuru-Guzik A. Expressibility and entangling capability of parameterized quantum circuits for hybrid quantum-classical algorithms. *Adv Quantum Tech* (2019) **2**:1900070. doi:10.1002/qute.201900070

25. Crawford O, Cruise JR, Mertig N, Gonzalez-Zalba MF. Compilation and scaling strategies for a silicon quantum processor with sparse two-dimensional connectivity. *npj Quantum Inf* (2023) **9**:13. doi:10.1038/s41534-023-00679-8

26. Hashim A, Rines R, Omole V, Naik RK, Kreikebaum JM, Santiago DI, Chong FT, Siddiqi I, Gokhale P. Optimized SWAP networks with equivalent circuit averaging for QAOA. *Phys Rev Res* (2022) **4**:033028. doi:10.1103/PhysRevResearch.4.033028

27. Kivlichan ID, McClean J, Wiebe N, Gidney C, Aspuru-Guzik A, Chan GK-L, Babbush R. Quantum Simulation of Electronic Structure with Linear Depth and Connectivity. *Phys Rev Lett* (2018) **120**:110501. doi:10.1103/PhysRevLett.120.110501

28. Yuan P, Zhang S. Full Characterization of the Depth Overhead for Quantum Circuit Compilation with Arbitrary Qubit Connectivity Constraint. *Quantum* (2025) **9**:1757. doi:10.22331/q-2025-05-28-1757

29. AbuGhanem M. IBM quantum computers: evolution, performance, and future directions. *J Supercomput* (2025) **81**:687. doi:10.1007/s11227-025-07047-7

30. Cross AW et al. Validating quantum computers using randomized model circuits. *Physical Review A* (2019) **100**:032328.

31. Javadi-Abhari A et al. Quantum computing with Qiskit. *arXiv preprint* (2024)

32. Team IQ. Qiskit: An open-source framework for quantum computing. (2024) Available at: https://qiskit.org/

33. Preskill J. Quantum Computing in the NISQ era and beyond. *Quantum* (2018) **2**:79. doi:10.22331/q-2018-08-06-79

34. Bravyi S, Kitaev A. Universal quantum computation with ideal Clifford gates and noisy ancillas. *Phys Rev A* (2005) **71**:022316. doi:10.1103/PhysRevA.71.022316

35. Farhi E, Goldstone J, Gutmann S. A Quantum Approximate Optimization Algorithm. (2014) doi:10.48550/arXiv.1411.4028

36. Garcez A d'Avila, Lamb LC. Neurosymbolic AI: the 3rd wave. *Artif Intell Rev* (2023) **56**:12387–12406. doi:10.1007/s10462-023-10448-w

37. Meyer DA. Quantum Strategies. *Phys Rev Lett* (1999) **82**:1052–1055. doi:10.1103/PhysRevLett.82.1052

38. Piotrowski EW, Sładkowski J. An Invitation to Quantum Game Theory. *International Journal of Theoretical Physics* (2003) **42**:1089–1099. doi:10.1023/A:1025443111388

# Supplementary Material

# Logical Consistency as a Dynamical Invariant:
# A Quantum Model of Self-Reference and Paradox

N. Cheimarios[1,*,**] and S. Cheimariou[2]

[1]School of Chemical Engineering, National Technical University of Athens

Iroon Polytechneiou 9, Zografou 157 72, Athens, Greece

[2] Department of Speech Pathology, Xavier University of Louisiana, 1 Drexel Dr, New Orleans, New Orleans, Louisiana, USA

*Corresponding author: nixeimar@chemeng.ntua.gr

**Current address: Accenture, 1 Arcadias 14564 Kifissia, Athens, Greece

**S1. Fixed-Point Logical Consistency Proof**

Let the total Hilbert space be,

$$\Pi = \otimes_{i=1}^{n} \mathcal{H}_{q_{c_i}} \otimes_{i=1}^{n} \mathcal{H}_{q_{r_i}} \otimes \mathcal{H}_{q_f} \tag{S1.1}$$

where $q_{c_i}$ denotes the contradiction qubit, $q_{r_i}$ the resolution qubit, and $q_f$ the global consistency flag. Each $\mathcal{H}$ is a two-dimensional Hilbert space spanned by $|0\rangle$ and $|1\rangle$.

For each pair $(q_{c_i}, q_{r_i})$, define the local logical constraint state $|q_{c_i} = 1, q_{r_i} = 0\rangle$, representing an unresolved contradiction. The local consistency projector is,

$$\Pi_i = \mathrm{I}_{c_i r_i} - |c_i = 1, r_i = 0\rangle\langle c_i = 1, r_i = 0| \tag{S1.2}$$

The global logical-consistency projector is,

$$\Pi = \bigotimes_i \Pi_i \tag{S1.3}$$

whose image Im($\Pi$) represents all configurations where every contradiction is resolved.

For each contradiction–resolution pair, define the Toffoli-type unitary

$$U = \text{Toffoli}(q_{c_i}, \neg q_{r_i}; f) \tag{S1.4}$$

which flips the global flag $f$ if and only if $q_{c_i} = 1$ and $q_{r_i} = 0$. The global unitary is the product

$$U = \prod_{i=1}^{n} U_i \tag{S1.5}$$

which is unitary since each component acts on distinct control registers.

**Proposition**

Let

$$\mathcal{F} = \{|\psi\rangle \in \mathcal{H} : U|\psi\rangle = |\psi\rangle\} \tag{S1.6}$$

denote the fixed-point subspace of $U$. Then,

$$\mathcal{F} = \text{Im}(\Pi) \otimes \text{span}\{|f = 1\rangle\} \tag{S1.7}$$

That is, $U$ acts as the identity on all states satisfying logical consistency and leaves the flag $f = 1$, while any inconsistent configuration exits this subspace.

**Proof**

Let $|c, r, f\rangle$ be a computational basis state.

(i) if for all $i$, $(c_i, r_i) \neq (1,0)$, then no Toffoli gate flips $f$, and

$$U|c, r, f\rangle = |c, r, f\rangle \tag{S1.8}$$

Hence all states in $\text{Im}(\Pi)\otimes |f=1\rangle$ are fixed points.

(ii) If there exists at least one $i$ such that $(c_i, r_i) = (1,0)$, then $U_i$ flips $f$,

$$U|c, r, f\rangle = |c, r, 0\rangle \tag{S1.9}$$

So the state is no longer invariant. Thus, no inconsistent state lies in the fixed-point subspace.

Therefore,

$$\mathcal{F} = \text{Im}(\Pi)\otimes\text{span}\{|f=1\rangle\}$$

**S2 Projector-Hamiltonian Equivalence and Fixed-Point Subspace**

To clarify the formal structure of the logical-coherence operator and its relation to known projector-based evolutions, we derive the explicit correspondence between the unitary operator $U = 2\Pi - I$ and the exponential form of a projector Hamiltonian. Let $\Pi$ denote the global consistency projector acting on the Hilbert space

$$\Pi = \otimes_{k=i}^{N} \mathbb{C}^2 \tag{S2.1}$$

of dimension $2^N$ with $\Pi^2 = \Pi = \Pi^\dagger$.

We define the corresponding logical Hamiltonian

$$H_{logic} = I - \Pi \tag{S2.2}$$

which penalizes logically inconsistent configurations (contradictory subspace). By construction, $H_{logic}$ is Hermitian and positive semidefinite, with eigenvalues $\{0,1\}$.

**S3.1 Unitary Exponential Form**

For any projector $P$ satisfying $P^2 = P$, the exponential map $e^{-i\theta P}$ acts as

$$e^{-i\theta P} = I + \left(e^{-i\theta P} - 1\right)P \tag{S2.3}$$

Setting $\theta = \pi$ gives,

$$e^{-i\pi P} = I - 2P \tag{S2.4}$$

Applying this to Eq. S2.2, we obtain,

$$e^{-i\pi H_{logic}} = e^{-i\pi(I-\Pi)} = e^{-i\pi}e^{-i\pi\Pi} = -I + 2\Pi = 2\Pi - I \tag{S2.5}$$

Hence,

$$U = 2\Pi - I = e^{-i\pi H_{logic}} \tag{S2.6}$$

This shows that the logical-coherence operator can be expressed exactly as the exponential of a projector Hamiltonian evaluated at rotation angle $\pi$. The equivalence ensures that $U$ is both Hermitian and unitary, with eigenvalues $\pm 1$. Thus, $U$ is a reflection about the subspace $\text{Im}(\Pi)$, not a dynamical evolution toward its ground state.

### S3.2 Fixed-Point–Kernel Equivalence

Let $|\psi\rangle \in \mathcal{H}$ be an arbitrary state. From Eqs. (S2.2)–(S2.6),

$$U|\psi\rangle = |\psi\rangle \iff (2\Pi - I)|\psi\rangle = |\psi\rangle \iff \Pi|\psi\rangle = |\psi\rangle \tag{S2.7}$$

Acting with $H_{logic}$ on such state yields

$$H_{logic}|\psi\rangle = (I - 2\Pi)|\psi\rangle = 0 \tag{S2.8}$$

Thus,

$$U|\psi\rangle = |\psi\rangle \iff H_{logic}|\psi\rangle = 0 \iff |\psi\rangle \in \ker(H_{logic}) \tag{S2.9}$$

Therefore, the logical-coherence subspace preserved by the circuit's unitary evolution is identical to the ground-state (kernel) subspace of $H_{logic}$ i.e. the fixed-point subspace of the unitary evolution coincides exactly with the kernel of the logical Hamiltonian, establishing a one-to-one correspondence between *dynamical invariance* and *logical consistency*.

Equations (S2.6)–(S2.9) demonstrate that $U$ formally belongs to the same algebraic family as reflections used in amplitude amplification (see [1]). However, the projector $\Pi$ here represents a logical-consistency condition, not a data-selection oracle. Therefore, the resulting unitary enforces consistency dynamically—through a single coherent reflection that eliminates contradictory amplitudes—rather than via iterative amplitude amplification or energetic minimization as in QCS/QSAT models. This connection situates the present construction within the general theory of projector-generated unitaries while highlighting its distinct operational semantics.

## S3. Raw data

**Table S1**: Average counts after 20 runs with 8192 shots

| State | Simulation | | Hardware | |
|---|---|---|---|---|
| | Counts | Probability | Counts | Probability |
| 0000 | 1 | 0.0001 | 1 | 0.0001 |
| 0001 | 34 | 0.0040 | 40 | 0.0046 |
| 0010 | 28 | 0.0033 | 34 | 0.0040 |
| 0011 | 1 | 0.0001 | 0 | 0.0000 |
| 0100 | 0 | 0.0000 | 1 | 0.0001 |
| 0101 | 2 | 0.0002 | 1 | 0.0001 |
| 0110 | 0 | 0.0000 | 3 | 0.0003 |
| 1000 | 249 | 0.0293 | 166 | 0.0191 |
| 1001 | 3460 | 0.4068 | 3680 | 0.4243 |
| 1010 | 3956 | 0.4645 | 3792 | 0.4371 |
| 1011 | 105 | 0.0123 | 68 | 0.0079 |
| 1100 | 44 | 0.0052 | 84 | 0.0097 |
| 1101 | 138 | 0.0162 | 140 | 0.0162 |

| | 1110 | 110 | 0.0129 | 122 | 0.0141 |
| | 1111 | 64 | 0.0075 | 60 | 0.0069 |

## S4 Asymptotic Complexity

Each Toffoli gate decomposes into six CNOTs and nine single-qubit rotations in the standard IBM basis, giving a gate depth of approximately

$$D(N) \approx 6\left(\frac{N}{2}\right) + D_{1q}(N) \sim O(N) \qquad (S4.1)$$

where $D_{1q}(N)$ accounts for the single-qubit rotation and barrier layers. The total two-qubit gate count therefore scales linearly,

$$G_{2q} = 3N \qquad (S4.2)$$

since each contradiction–resolution pair contributes one Toffoli (6 CNOTs) and one controlled-phase (1 CNOT-equivalent depth). This linear scaling is notably lighter than the quadratic scaling typical of QSAT or variational-constraint circuits, where all-to-all connectivity is required [2, 3].

## S5 Connectivity Constraints

On IBM backends such as *ibm_torino*, qubits are arranged in a heavy-hex lattice with at most three nearest neighbors. Because each Toffoli gate involves two controls and one target, SWAP operations must be inserted whenever the controls are not directly connected. If the logical qubits are mapped linearly, the average routing overhead contributes an additional term

$$D_{SWAP}(N) \approx 2(mean\ control - target\ distance) \sim O(N) \qquad (S5.1)$$

where the factor of 2 accounts for the forward and backward SWAP operations required to route qubits across limited-connectivity hardware. This linear dependence arises because the average control–target distance typically grows proportionally with the number of qubits $N$ on a one-dimensional or sparse coupling graph [4, 5], so the total practical depth remains linear,

The total circuit depth therefore scales as [6, 7],

$$D_{total}(N) \sim O(N) + O(N) = O(N) \tag{S5.2}$$

with a constant factor increase for unoptimized routing.

## S6 Error Accumulation

Let $p_{2q}$ and $p_{1q}$ denote the typical two-qubit and single-qubit error probabilities. The expected fidelity for a single circuit execution is approximated by

$$F_N \approx (1-p_{2q})^{G_{2q}(N)}(1-p_{1q})^{G_{1q}(N)} \approx e^{\left[-\left(p_{2q}G_{2q}(N)+p_{2q}G_{2q}(N)\right)\right]} \tag{S6.1}$$

Using published calibration numbers for current-generation processors (for example, median two-qubit gate error rates approaching about $10^{-3}$ on some IBM backends, and single–qubit error rates below $10^{-4}$) [8].

If one uses $p_{2q} \approx 10^{-3}$ and $p_{1q} \approx 10^{-4}$, and substitutes for a circuit with $G_{2q}(N) \approx 3N$ (and neglects the much smaller single-qubit contribution), then one finds approximately

$$F_N \approx e^{[-(-0.003\ N)]} \tag{S6.2x}$$

This suggests that for such error rates, maintaining high fidelity becomes increasingly difficult once $N$ grows beyond $O(10)$ qubits (since $e^{[-(-0.003\ N)]}$ drops significantly). In other words, on NISQ devices, cumulative two-qubit errors will dominate and limit circuit size unless error rates or gate counts are improved.

## References

1. Nielsen, M.A., Chuang, I.L.: Quantum Computation and Quantum Information: 10th Anniversary Edition, https://www.cambridge.org/highereducation/books/quantum-computation-and-quantum-information/01E10196D0A682A6AEFFEA52D53BE9AE

2. Yuan, P., Allcock, J., Zhang, S.: Does qubit connectivity impact quantum circuit complexity?, http://arxiv.org/abs/2211.05413, (2023)
3. Sim, S., Johnson, P.D., Aspuru-Guzik, A.: Expressibility and entangling capability of parameterized quantum circuits for hybrid quantum-classical algorithms. Adv Quantum Tech. 2, 1900070 (2019). https://doi.org/10.1002/qute.201900070
4. Crawford, O., Cruise, J.R., Mertig, N., Gonzalez-Zalba, M.F.: Compilation and scaling strategies for a silicon quantum processor with sparse two-dimensional connectivity. npj Quantum Inf. 9, 13 (2023). https://doi.org/10.1038/s41534-023-00679-8
5. Hashim, A., Rines, R., Omole, V., Naik, R.K., Kreikebaum, J.M., Santiago, D.I., Chong, F.T., Siddiqi, I., Gokhale, P.: Optimized SWAP networks with equivalent circuit averaging for QAOA. Phys. Rev. Res. 4, 033028 (2022). https://doi.org/10.1103/PhysRevResearch.4.033028
6. Kivlichan, I.D., McClean, J., Wiebe, N., Gidney, C., Aspuru-Guzik, A., Chan, G.K.-L., Babbush, R.: Quantum Simulation of Electronic Structure with Linear Depth and Connectivity. Phys. Rev. Lett. 120, 110501 (2018). https://doi.org/10.1103/PhysRevLett.120.110501
7. Yuan, P., Zhang, S.: Full Characterization of the Depth Overhead for Quantum Circuit Compilation with Arbitrary Qubit Connectivity Constraint. Quantum. 9, 1757 (2025). https://doi.org/10.22331/q-2025-05-28-1757
8. AbuGhanem, M.: IBM quantum computers: evolution, performance, and future directions. J Supercomput. 81, 687 (2025). https://doi.org/10.1007/s11227-025-07047-7